\documentclass[aps, prd, amsmath, floats, floatfix, twocolumn, nofootinbib, nosuperscriptaddress]{revtex4-1}
\usepackage[pdftex]{graphicx}
\usepackage{color}
\usepackage{latexsym}
\usepackage[colorlinks]{hyperref}
\usepackage{amsthm,amsmath,amssymb}
\usepackage{mathrsfs}
\usepackage{makecell}

\newcommand{\be}{\begin{eqnarray}}
\newcommand{\ee}{\end{eqnarray}}

\begin{document}

\title{Testing General Relativity with Black Holes}

\author{Cosimo~Bambi}
\email[]{bambi@fudan.edu.cn}
\affiliation{Center for Astronomy and Astrophysics, Center for Field Theory and Particle Physics, and Department of Physics,
Fudan University, Shanghai 200438, China}
\affiliation{School of Natural Sciences and Humanities, New Uzbekistan University, Tashkent 100007, Uzbekistan}

\begin{abstract}
The theory of General Relativity has successfully passed a large number of observational tests. The theory has been extensively tested in the weak-field regime with experiments in the Solar System and observations of binary pulsars. The past 10~years have seen significant advancements in the study of the strong-field regime, which can now be tested with gravitational waves, X-ray data, and black hole imaging. Here I summarize the state-of-the-art of the tests of General Relativity with black hole X-ray data and I briefly discuss the long-term vision of the possibility of an interstellar mission to the closest black hole for more precise and accurate tests.
\end{abstract}

\maketitle

%%%%%%%%%%%%%%%%%%%%%%%%%%%%%%

\section{Introduction}

Black holes are the sources of the strongest gravitational fields in the Universe today and can be used as laboratories for testing General Relativity in the strong-field regime~\cite{Bambi:2017khi}. The past 10~years have significantly changed this line of research. Until 2015, we knew that stellar-mass black holes in X-ray binary systems were dark and compact objects with a mass exceeding the maximum mass for a neutron star~\cite{Rhoades:1974fn,Casares:2013tpa} and that supermassive black holes in galactic nuclei were too massive, compact, and old to be clusters of neutron stars~\cite{Maoz:1997yd}. There was no evidence that these objects were the Kerr black holes predicted by General Relativity: the latter was just the simplest interpretation and the only one that did not require new physics~\cite{Bambi:2015kza}. The situation could be potentially similar to the violation of the parity symmetry in weak interactions: there were no experimental tests of the conservation of parity in weak interactions before 1956 and the experiment by Chien-Shiung Wu in 1956 found that the conservation of parity was violated by weak interactions~\cite{Wu:1957my}. 

In the past 10~years, we have learnt to test the nature of black holes with gravitational waves~\cite{LIGOScientific:2016lio,Yunes:2016jcc,LIGOScientific:2019fpa,LIGOScientific:2020tif,Shashank:2021giy}, X-ray data~\cite{Cao:2017kdq,Tripathi:2018lhx,Tripathi:2019bya,Tripathi:2020dni,Tripathi:2020yts,Zhang:2021ymo,Tripathi:2021rqs}, and black hole imaging~\cite{Bambi:2008jg,Bambi:2019tjh,EventHorizonTelescope:2020qrl,Vagnozzi:2022moj,EventHorizonTelescope:2022xqj}. Current constraints may not be considered very stringent, especially if compared to our capability of testing atomic and particle physics, but we are just at the beginning and these constraints can be improved with future observational facilities and more advanced theoretical models.    

In what follows, I will try to summarize the state-of-the-art of the tests of black holes with X-ray data. As of now, these tests provide very competitive constraints on the Kerr hypothesis, namely that astrophysical black holes are the Kerr black holes predicted by General Relativity. A part of the scientific community may be still a bit skeptical about the accuracy of these tests, but in the past years there have been significant efforts to understand better the systematic uncertainties of X-ray techniques and to develop more advanced theoretical models. Moreover, unlike population studies in astrophysics, where we want to study the properties of as many sources as possible, in the case of tests of the Kerr hypothesis we can focus our studies on a very limited number of sources and observations for which we have very high-quality data, the astrophysical environment is understood well, and systematic uncertainties are under control.   

Current observational tests of black holes will be improved by the next generation of observational facilities. However, at some point the constraining capability of our astrophysical observations will be likely limited by complications of the astrophysical environment and simplifications in our theoretical models. In general, we may not be able to perform very precise and accurate tests of the nature of astrophysical black holes with distant observations of these objects. In this regards, we might consider the possibility of an interstellar mission to send a small spacecraft to the nearest black hole in order to obtain information about black holes and General Relativity that might be difficult to obtain in other ways.

%%%%%%%%%%%%%%%

\begin{figure*}[t]
\centering
\includegraphics[width=0.45\linewidth]{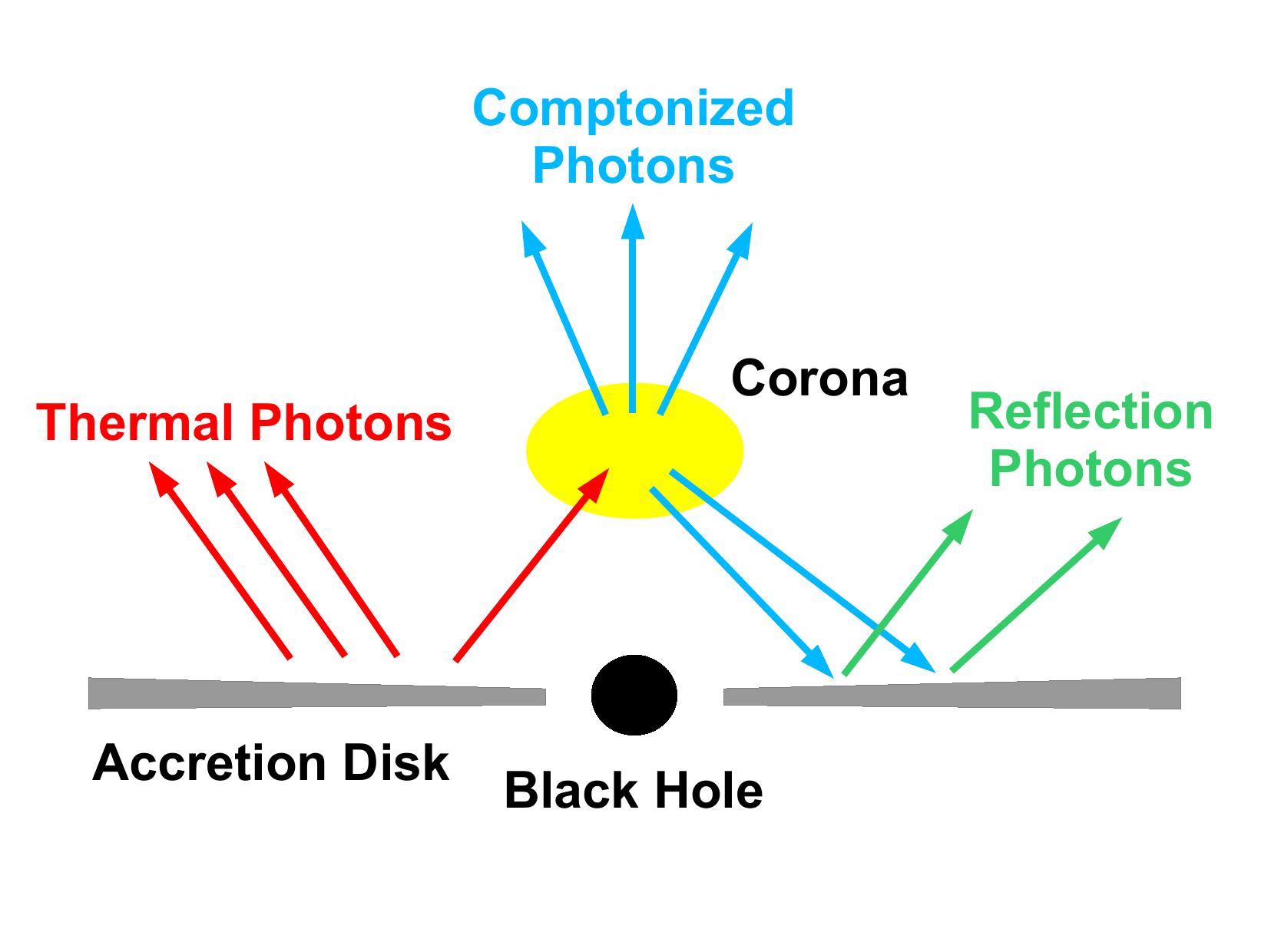}
\hspace{0.5cm}
\includegraphics[width=0.45\linewidth]{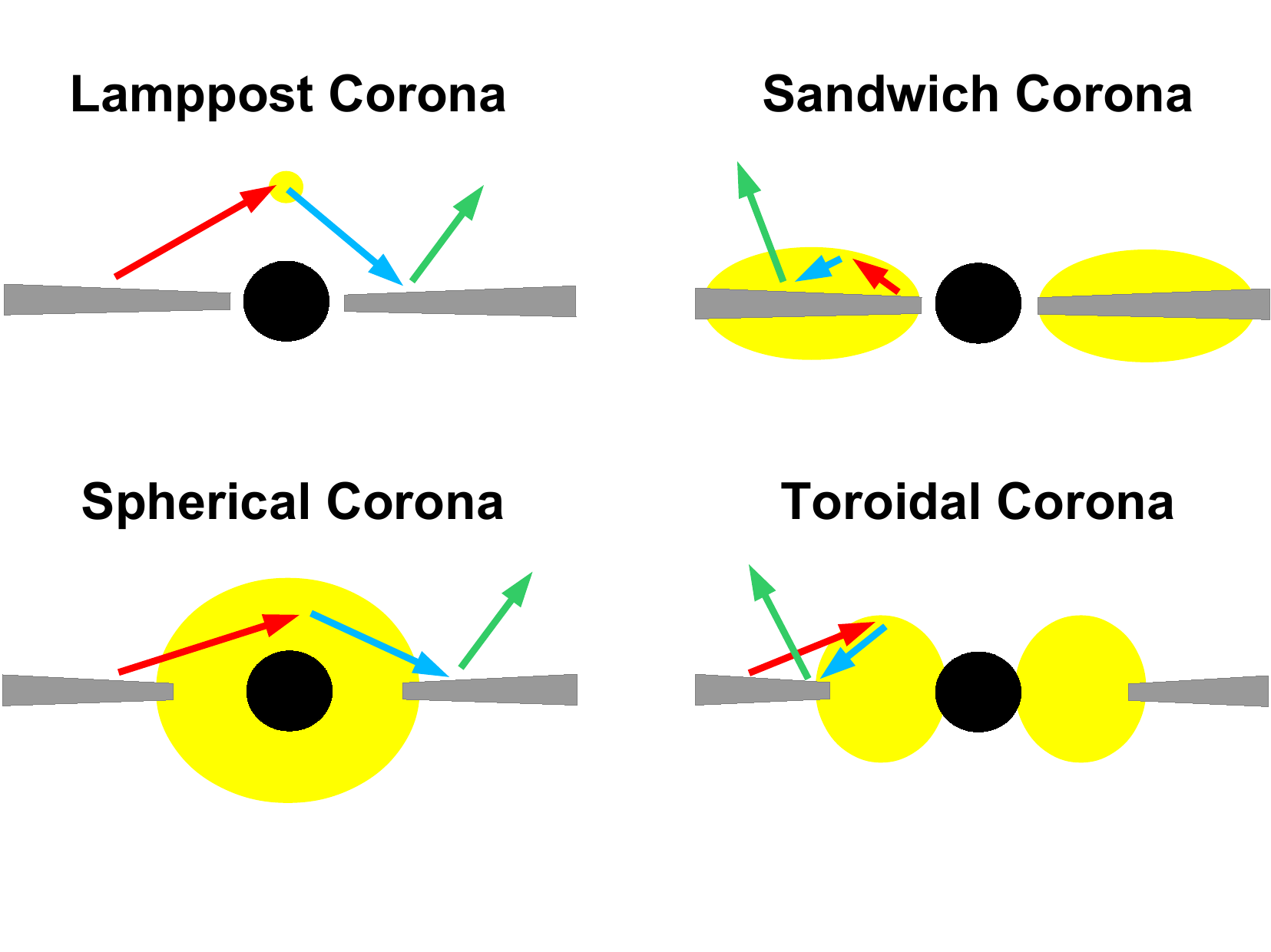}
\vspace{-0.8cm}
\caption{Left panel: Disk-corona model. Right panel: Examples of possible coronal geometries. Figure from Ref.~\cite{Bambi:2024hhi}.}
\label{f-corona}
\end{figure*}

\section{Testing Black Holes with X-Ray Data}

\subsection{Disk-Corona Model}

The astrophysical system that we can study with X-ray observations is shown in the left panel of Fig.~\ref{f-corona} and is normally referred to as the {\it disk-corona model}~\cite{Bambi:2020jpe,Bambi:2024hhi}. The black hole can be either a stellar-mass black hole in an X-ray binary system or a supermassive black hole in an active galactic nucleus. The crucial ingredient is that the accretion disk is geometrically thin and optically thick and that the motion of the material of the disk is Keplerian. Such an accretion disk is {\it cold} (with respect to other accretion disk models, like an advection-dominated accretion flow~\cite{Narayan:1994xi}). The thermal spectrum of the accretion disk is normally peaked in the soft X-ray band (0.1-10~keV) in the case of stellar-mass black holes in X-ray binary systems and in the UV band (1-100~eV) in the case of supermassive black holes in active galactic nuclei~\cite{Shakura:1972te,Page:1974he}. The ``corona'' is some {\it hot} plasma (with an electron temperature $T_{\rm e} \sim 100$~keV) near the black hole and the inner part of the accretion disk, but its exact nature and geometry are not yet well understood. Possible coronal geometries are shown in the right panel of Fig.~\ref{f-corona}. In the lamppost model, the corona is compact and along the black hole spin axis; for example, the base of the jet may act as a lamppost corona~\cite{Markoff:2005ht}. In the sandwich model, the corona is a layer covering the accretion disk; the hot atmosphere of the accretion disk may act as a sandwich corona~\cite{Titarchuk:1994rz,Petrucci:2012cv}. If the accretion disk is truncated at large radii, we may have an advection-dominated accretion flow between the inner edge of the cold disk and the black hole; such an advection-dominated accretion flow may act as a spherical or toroidal corona~\cite{Dove:1997ei}.

Since the disk is cold and the corona is hot, thermal photons of the disk can inverse Compton scatter off free electrons in the corona. The spectrum of the Comptonized photons can be approximated by a power-law with a high-energy cutoff (determined by the electron temperature of the corona) and a low-energy cutoff (determined by the temperature of the accretion disk)~\cite{Zdziarski:1996wq,Zdziarski:2019cvs}. A fraction of the Comptonized photons can illuminate the disk: Compton scattering and absorption followed by fluorescent emission generate the reflection spectrum (green arrows in Fig.~\ref{f-corona}). The {\it non-relativistic reflection spectrum}, namely the reflection spectrum in the rest-frame of the material of the disk, is characterized by narrow fluorescent emission lines in the soft X-ray band and a Compton hump peaking around 20-50~keV~\cite{Ross:2005dm,Garcia:2010iz}. The {\it relativistic reflection spectrum}, namely the reflection spectrum of the whole disk as observed far from the source, is blurred due to relativistic effects: Doppler boosting because of the motion of the material in the disk and gravitational redshift because of the gravitational well of the black hole~\cite{Bambi:2020jpe,Bambi:2024hhi}.  

The {\it continuum-fitting method} refers to the analysis of the thermal spectrum of Keplerian, geometrically thin, and optically thick accretion disks around black holes and has been originally developed for measuring black hole spins (under the assumption that the spacetime geometry is described by the Kerr solution)~\cite{Zhang:1997dy,Li:2004aq,McClintock:2011zq,McClintock:2013vwa}. This technique is normally used only for stellar-mass black holes in X-ray binary systems. In the case of supermassive black holes in active galactic nuclei, the thermal spectrum of the disk is peaked in the UV band, where dust absorption limits the possibility of accurate measurements of the thermal spectrum of a source. For a Keplerian and infinitesimally thin accretion disk, the time-averaged radial structure of the disk directly follows from the conservation of mass, energy, and angular momentum~\cite{Page:1974he}. If we assume the Kerr hypothesis, the model depends on 5~parameters: the black hole mass, the black hole spin angular momentum, the mass accretion rate, the inclination angle of the disk with respect to the line of sight of the observer, and the distance between the source and the observer. If we can get independent estimates of the black hole mass, the inclination angle of the disk, and the distance of the source (for example, from optical observations of the companion star~\cite{Casares:2013tpa}), we can fit the thermal spectrum of the disk and infer the black hole spin angular momentum and the mass accretion rate.

{\it X-ray reflection spectroscopy} refers to the analysis of the relativistically blurred reflection features in the X-ray spectra of accreting black holes. This technique can be used to measure spins of both stellar-mass black holes in X-ray binary systems and supermassive black holes in active galactic nuclei~\cite{Brenneman:2006hw,Bambi:2020jpe,Bambi:2024hhi,Draghis:2022ngm,Draghis:2025izq}. It has the advantages over the continuum-fitting method that it does not require independent measurements of the black hole mass, inclination angle of the disk, and distance of the source. The reflection spectrum indeed does not directly depend on the black hole mass and distance, and the inclination angle of the disk can be inferred from the analysis of the reflection features itself.

\subsection{Testing New Physics}

If we want to calculate the thermal spectrum of a thin accretion disk around a black hole, we need an astrophysical model for the accretion disk and we have to know the motion of massive and massless particles around the black hole. In the case of the reflection spectrum, we need even an astrophysical model for the corona and we have to know atomic physics near the black hole. If we have the astrophysical part under control (models for the disk and for the corona), we can test the motion of massive and massless particles around the black hole and atomic physics in a strong gravitational field. 

In General Relativity and in the absence of exotic matter field, we expect that the spacetime geometry around astrophysical black holes is approximated well by the Kerr solution (Kerr hypothesis); see, for example, Refs.~\cite{Bambi:2017khi,Bambi:2014koa,Bambi:2008hp}. However, there are models beyond General Relativity where the final product of gravitational collapse is not a Kerr black hole. We can use the continuum-fitting method and X-ray reflection spectroscopy to test the Kerr hypothesis~\cite{Cao:2017kdq,Tripathi:2018lhx,Tripathi:2019bya,Tripathi:2020dni,Tripathi:2020yts,Zhang:2021ymo,Tripathi:2021rqs,Tripathi:2020qco}. These two techniques can also test the Weak Equivalence Principle in the strong gravitational fields of black holes~\cite{Roy:2021pns}. In General Relativity, all free particles follow the geodesics of the spacetime and therefore their motion is independent of their internal structure and composition (Weak Equivalence Principle). However, there are models beyond General Relativity where this is not true: for example, because of non-universal interactions between the gravity and the matter sectors, protons and electrons in the accretion disk may follow the geodesics of the spacetime and photons may not (or vice versa). Last, in General Relativity and in any metric theory of gravity, the atomic physics near a black hole must be the same as in our laboratories on Earth: this is because in all these models gravity universally couples to matter and locally all non-gravitational laws of physics reduce to those of Special Relativity. Again, this is not true in some extensions of General Relativity, where we can have phenomena like the variation of fundamental constants~\cite{Bambi:2013mha}. 

In the past years, my group at Fudan University has developed the models {\tt relxill\_nk}~\cite{Bambi:2016sac,Abdikamalov:2019yrr,Abdikamalov:2020oci} and {\tt nkbb}~\cite{Zhou:2019fcg}, which are both public on GitHub\footnote{\href{https://github.com/ABHModels}{https://github.com/ABHModels}}. {\tt relxill\_nk} is an extension of the {\tt relxill} package~\cite{Dauser:2013xv,Garcia:2013oma,Garcia:2013lxa} for non-Kerr spacetimes: it is a reflection model specifically designed to test the Kerr hypothesis. {\tt nkbb} is a thermal model to test the Kerr hypothesis. These models have been extensively used to test the nature of stellar-mass black holes in X-ray binaries and supermassive black holes in active galactic nuclei. The results are briefly summarized in the next subsection.

\subsection{Results}

If we want to test the Kerr nature of astrophysical black holes, we can follow either an {\it agnostic} (or bottom-up) approach or a {\it theory-specific} (or top-down) method. With an agnostic approach, we want to constrain/measure generic deviations from the Kerr geometry around a source~\cite{Zhao:2023erf}. With a theory-specific method, we want to test General Relativity against another theory of gravity in which the final product of the gravitational collapse is not described by the Kerr solution. 

\begin{figure*}[t]
\centering
\includegraphics[width=0.7\linewidth]{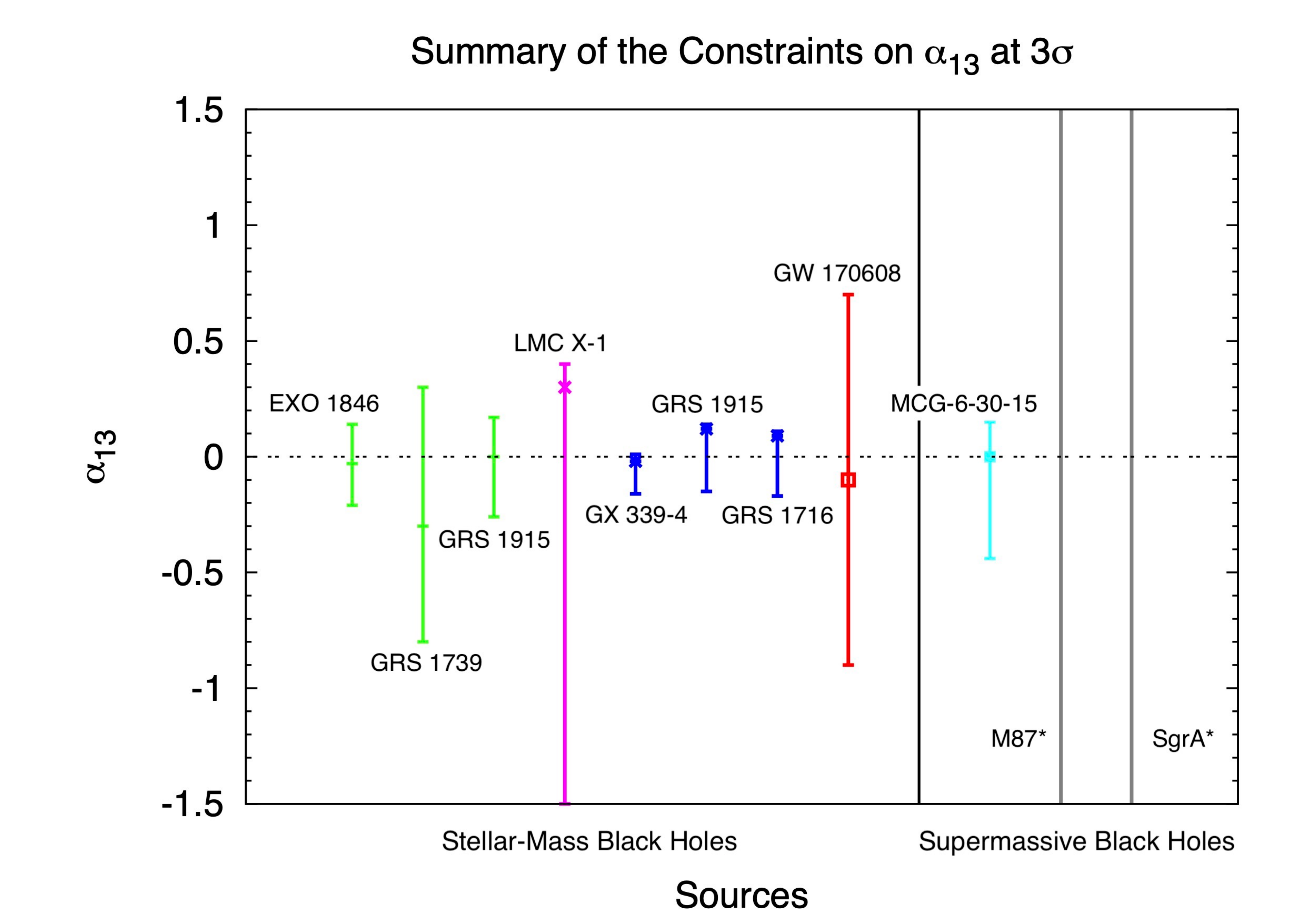}
%\vspace{-0.8cm}
\caption{Summary of current 3-$\sigma$ constraints on the Johannsen deformation parameter $\alpha_{13}$ with X-ray reflection spectroscopy (error bars in green for stellar-mass black holes and error bar in cyan for supermassive black holes), continuum-fitting method (error bar in magenta), combination of X-ray reflection spectroscopy and continuum-fitting method (error bars in blue), gravitational waves (error bar in red), and black hole imaging (error bars in gray). See the text and Ref.~\cite{Bambi:2022dtw} for more details.}
\label{f-summary}
\end{figure*}

Fig.~\ref{f-summary} summarizes current constraints within an agnostic approach. We assume that the spacetime geometry around black holes is described by the Johannsen metric with a possible non-vanishing deformation parameter $\alpha_{13}$~\cite{Johannsen:2013szh} and we try to measure such a deformation parameter to check if it vanishes, which is the condition to recover the Kerr metric. The error bars in green are the three most robust and precise measurements of $\alpha_{13}$ with X-ray reflection spectroscopy for stellar-mass black holes in X-ray binaries~\cite{Tripathi:2020yts}. The error bar in magenta is the constraint on $\alpha_{13}$ with the continuum-fitting method for the stellar-mass black hole in LMC~X-1: the constraint is weak because the thermal spectrum has a simple shape and there is a strong degeneracy between the estimate of the black hole spin and the estimate of $\alpha_{13}$~\cite{Tripathi:2020qco}. For some sources, we can combine the analysis of the thermal spectrum with that of the reflection features and get more stringent and robust constraints on $\alpha_{13}$: this is the case of the error bars in blue and the three sources are the X-ray binaries GX~339--4~\cite{Tripathi:2020dni}, GRS~1915+105~\cite{Tripathi:2021rqs}, and GRS~1716--249~\cite{Zhang:2021ymo}. The error bar in red is the most stringent constraint on $\alpha_{13}$ from gravitational wave data (from the Gravitational Wave Transient Catalog 3, GWTC-3): it is obtained assuming that the spacetime geometry is described by the Johannsen metric and the gravitational wave emission is described by the Einstein equations (see Refs~\cite{Shashank:2021giy,Das:2024mjq} for more details). The error bar in cyan is the most robust and precise measurement of $\alpha_{13}$ with X-ray reflection spectroscopy for supermassive black holes~\cite{Tripathi:2018lhx}. The (very weak) constraints in gray are current results from black hole imaging with the Event Horizon Telescope~\cite{EventHorizonTelescope:2020qrl,EventHorizonTelescope:2022xqj}. 

{\tt relxill\_nk} has been used even to test specific gravity theories in which uncharged black holes are not described by the Kerr solution: conformal gravity~\cite{Zhou:2018bxk}, Kaluza-Klein gravity~\cite{Zhu:2020cfn}, asymptotically safe quantum gravity~\cite{Zhou:2020eth}, Einstein-Maxwell-dilaton-axion gravity~\cite{Tripathi:2021rwb}.

All current measurements are consistent with the Kerr hypothesis of General Relativity. X-ray tests provide very competitive constraints, definitively more stringent than those from black hole imaging with current data of the Event Horizon Telescope. X-ray tests can also be seen as complementary to gravitational wave tests, even in the future: generally speaking, gravitational waves should be more suitable to test the gravity sector and can directly probe the Einstein equations, while X-ray observations should be more suitable to test the interactions between the matter and the gravity sectors. For example, new interaction terms between the matter and the gravity sectors may lead to non-geodesic motion of certain particles, variation of fundamental constants, etc. All these phenomena can leave a signature in the black hole electromagnetic spectra without affecting the gravitational wave signal. 

X-ray tests of the Kerr hypothesis can be both precise and accurate if we properly select and sources and the observations to analyze. X-ray reflection spectroscopy provides better results when most of the reflection radiation comes from the region very close to the black hole, which requires that the source is a fast-rotating black hole and that the corona is compact and very close to the black hole. For such sources, the reflection spectrum can have a very broadened iron line. It is also important to select sources with thin accretion disks, which requires that the Eddington-scaled accretion luminosity does not exceed 0.3. A detailed discussion on the criteria to select the sources for tests of the Kerr hypothesis and on the systematic uncertainties in these measurements can be found in Ref.~\cite{Bambi:2022dtw}.

%%%%%%%%%%%%%%%%%%%%%%%%%%%%%%%

\section{An interstellar mission to the closest black hole?}

The idea to send a small probe to the closest black hole to perform precise and accurate tests of the Kerr hypothesis was proposed in Ref.~\cite{Bambi:2025kcr} (see also Ref.~\cite{Bambi:2025hjn}). Although very speculative and extremely challenging, such a possibility is not completely unrealistic. Certainly we do not have the necessary technology today, but it may be available in 20-30~years. 

Stellar-mass black holes are the natural product of the evolution of heavy stars ($M_{\rm star} \gtrsim 20$~$M_\odot$); for a review, see Ref.~\cite{Bambi:2025rod}. Theoretical models predict $10^8$-$10^9$~stellar-mass black holes formed from the collapse of heavy stars in our Galaxy~\cite{Timmes:1995kp,Olejak:2019pln}. As of now, the closest {\it known} black hole is GAIA-BH1 at about 480~pc (1,560~light-years) from Earth~\cite{El-Badry:2022zih,Chakrabarti:2022eyq}. However, we can expect that there are many {\it unknown} black holes closer to us. It is impossible to make clear predictions and there are large uncertainties in current theoretical models. We should have $\sim 10^{10}$ white dwarfs in our Galaxy. If we believe that there are around $10^9$~stellar-mass black holes in our Galaxy~\cite{Timmes:1995kp}, we have roughly 1~black hole every 10~white dwarfs and we might expect that the closest black hole is at $\sim 16$~light-years from us\footnote{We know 5~white dwarfs within $\sim 16$~light-years of Earth.}. If we assume $10^8$~stellar-mass black holes in our Galaxy~\cite{Olejak:2019pln}, we have roughly 1~black hole every 100~white dwarfs and the closest black hole might be at $\sim 36$~light-years from us\footnote{We know 30~white dwarfs within $\sim 30$~light-years of Earth, so we can expect around 50~white dwarfs within $\sim 36$~light-years.}. 

Most stellar-mass black holes in our Galaxy are expected to be isolated black holes, without a companion star~\cite{Olejak:2019pln}. Discovering nearby isolated black holes is extremely challenging. The most promising strategy is to look for the electromagnetic radiation generated by an isolated black hole accreting from the interstellar medium~\cite{Meszaros75,McDowell85,Campana93,Fujita:1997fh,Tsuna:2018abi,Kimura:2021ayq,Murchikova:2025oio}. Murchikova \& Sahu (2025)~\cite{Murchikova:2025oio} pointed out that observational facilities like the Square Kilometer Array (SKA), the Atacama Large Millimiter/Submillimiter Array (ALMA), and James Webb Space Telescope (JWST) have the capability of detecting isolated stellar-mass black holes accreting from a warm interstellar medium within 150~light-years of Earth. It is possible that some nearby isolated black holes are already present in existing catalogs but they have not been identified as black holes. With a single telescope, an isolated black hole is detected as a faint source with a relatively featureless spectrum. Only if we observe such a source at different wavelengths we can realize that its spectrum is that expected from an isolated black hole accreting from the interstellar medium.

Let us assume we discover an isolated black hole that is not too far from us, like at 20~light-years (if we are more optimistic) or 40~light-years (if we are less optimistic). If we want to send a probe to such a black hole, it is clear that the probe should be able to travel at some fractions of the speed of light. This is impossible with traditional (chemically propelled) spacecrafts\footnote{The Tsiolkovsky rocket equation reads $m_i/m_f = e^{\Delta v / v_e}$, where $m_i$ is the initial total mass of the rocket (including propellant), $m_f$ is the final total mass of the rocket (without propellant), $\Delta v$ is the total change of the rocket's velocity, and $v_e$ is the effective exhaust velocity. If we use liquid hydrogen/liquid oxygen, we can have $v_e \sim 4.5$~km/s. If we want the rocket to travel at 1/10~of the speed of light, $m_i/m_f \sim 10^{28953}$; that is, even if $m_f$ were the proton mass, $m_i$ would exceed largely the total mass of the visible Universe!}. A solution could be the use of nanocrafts~\cite{Lubin16,Lubin22,Kuhlmey25}. A nanocraft is a gram-scale spacecraft. It has two main parts: a gram-scale wafer and a light sail. The gram-scale wafer is the main body of the satellite, with a computer processor, solar panels, navigation and communication equipment, etc. The light sail is an extremely thin, meter-scale, dielectric metamaterial used to accelerate the probe. Ground-based high-power lasers can accelerate the nanocraft through the radiation pressure of their laser beams on the light sail. After the nanocraft has reached its target velocity, the lasers are turned off. There are no specific technical problems to reach 90\% of the speed of light, but the cost of the laser is the most expensive part of the mission and higher target velocities can significantly increase the total cost of the mission.

If we have a black hole at 20~light-years from the Solar System, a nanocraft traveling at 1/3 of the speed of light can reach the object in 60~years, perform all experiments, and then we need 20~more years to receive the data on Earth. The mission would last around 80~years. If the black hole is at 40~light-years from the Solar System, we have to accelerate the nanocraft to a velocity closer to the speed of light if we want to remain with a mission under 100~years. With current technology, the cost of the lasers to accelerate a nanocraft at 1/3 of the speed of light would exceed one trillion EUR, which is definitively beyond the budget of every scientific experiment. We thus need to develop the technology to significantly reduce the cost of the lasers. If we consider the trend of the laser power and its price over the past 20~years and we extrapolate this trend to the future, we can expect that in 20-30~years the cost of the lasers to accelerate such a nanocraft will be around one billion EUR, which is compatible with the budget of a large space mission today.

When the nanocraft reaches the black hole, it should start its scientific experiment to tests the gravitational field around the compact object. If the nanocraft could orbit around the black hole, it could release a smaller probe and the latter could move to an orbit closer to the black hole: the nanocraft could communicate with the smaller probe by exchanging electromagnetic signals, which would allow us to study the motion of the smaller probe near the black hole and determine the nature of the compact object. The nanocraft could then send all data to Earth. If it is not possible to make the nanocraft orbit around the black hole, we should find a way to test the nature of the compact object when the nanocradt approaches the black hole. For example, when the nanocraft is approaching the black hole it could release a number of smaller probes and map the gravitational field around the compact object from the trajectories of these smaller probes. The nanocraft should then send all data to Earth.

%%%%%%%%%%%%%%%%%%%%%%%%%%%%%%%

\section{Concluding remarks}

General Relativity is one of the pillars of modern physics. For decades, the theory was extensively tested in the weak field regime with experiments in the Solar System and observations of binary pulsars, while the strong field regime was almost completely unexplored. Up to 10~years ago, we knew dark and compact objects in X-ray binaries with a mass exceeding the maximum mass for neutron stars and large amount of masses in relatively small volumes at the centers of many galaxies. We interpreted both object classes as the black holes predicted by General Relativity, but simply because it was the only interpretation that did not require new physics. There was no evidence that these objects were Kerr black holes. In the past 10~years, we have learnt to test the nature of these objects with gravitational waves, X-ray observations, and black hole imaging. All available observations have confirmed that these objects are the Kerr black holes predicted by General Relativity. Current constraints may not be very stringent, but they will be improved with the next generation of observational facilities. 

Here I briefly reviewed the state-of-the-art of the tests of General Relativity with black hole X-ray data. At the moment, we can test the Kerr hypothesis with the spectral analyses of the thermal and reflection components. In the future, we should be able to include even the polarization analysis of a source, which could provide independent constraints on possible deviations from the Kerr geometry and should help us to understand the geometry of the corona~\cite{Zhang:2025iae}.

I also presented a very speculative idea concerning the possibility of sending a small probe to a nearby black hole in order to perform very precise and accurate tests of the Kerr hypothesis. At the moment, we cannot say if a similar mission is feasible or unfeasible. Certainly we do not have the technology today, but it may be available in 20-30~years. It is also not obvious that a similar mission can test the Kerr hypothesis much better than astrophysical observations: it is possible, but accurate studies are necessary. In any case, the first step would be to find a close black hole. If such a black hole is within 20-25~light years of the Solar System, the necessary technology can be probably developed in a few decades. If the black hole is not within 20-25~light years, but still within 40-50~light years of the Solar System, the technological requirements for such a mission would be much more challenging but not out of reach. If there is no black hole within 40-50~light years of the Solar System, it is not a technological problem and every putative mission would last more than a century, which would make the project not very appealing.

%%%%%%%%%%%%%%%%%%%%%%%%%%%%%%%

\section*{Competing interests}
The author declares no competing interests.

\section*{Funding}
This work was supported by the National Natural Science Foundation of China (NSFC), Grant Nos.~W2531002, 12261131497, and 12250610185.

\section*{Data Availability}
This study did not analyze data.

%%%%%%%%%%%%%%%%%%%%%%%%%%%%%%%

\end{document}